\documentstyle[12pt]{article}
\newcommand{\beq}{\begin{equation}}
\newcommand{\eeq}{\end{equation}}
\begin{document}

\begin{center}
{\bf On the compatibility of causality and symmetry\\
(Comments on "Analysis of causality in time-dependent density
functional theory")} \bigskip \bigskip

M. Ya. Amusia$^{a,b}$ and V. R. Shaginyan$^{c,}$
\footnote{E--mail: vrshag@thd.pnpi.spb.ru}\bigskip

$^{a}$The Racah Institute of Physics, The Hebrew University,
Jerusalem 91904, Israel;\\[0pt] $^{b}$A.F. Ioffe
Physical-Technical Institute, St.Petersburg 194021, Russia;
\\[0pt] $^{c}$Petersburg  Nuclear Physics Institute,
Gatchina 188350, Russia\bigskip
\end{center}

\begin{abstract}
It is argued that there exists the only one inverse of the linear
response function $\chi$,  i.e. $\chi^{-1}$,  which depends
symmetrically of its spatial-times variables \cite{hb}.  Some
brief comments on this consideration are presented. We show
instead, that it is possible to construct the causal inverse
also. At the same time we confirm the main statement of \cite{hb}
that in fact there is no contradiction between the symmetry and
causality.\end{abstract}

\noindent PACS number(s): 31.15.Ew; 31.50.+w\bigskip

In a recent paper M. K. Harbola and A. Banerjee \cite {hb}
considered a paradox, which is connected to the symmetry
properties of the linear response function $\chi$ and the
causality (see, e.g. \cite{gdp,bg,l}). This paradox comes from an
observation, that the inverse of the linear response function of a
many-body system $\chi$, i.e. $\chi^{-1}({\bf r},t;{\bf
r}^{\prime},t^{\prime})$ being the second functional derivative,
\beq \chi^{-1}({\bf r},t;{\bf r}^{\prime},t^{\prime})
=\frac{\delta v_{ext}({\bf r},t)} {\delta\rho({\bf
r}^{\prime},t^{\prime})} =\frac{\delta^2 A[\rho]} {\delta\rho({\bf
r},t)\delta\rho({\bf r}^{\prime},t^{\prime})},\eeq is a symmetrical
function of their time and space variables. In eq. (1) $\rho({\bf
r},t)$ is the particle density of the system under consideration,
characterized by its action functional $A[\rho]$, and $v_{ext}$ is
an external field. On the other hand, one may believe that this
function must be a causal function, just as the linear response
function $\chi({\bf r},t;{\bf r}^{\prime},t^{\prime})$ itself,
which defines the variation of the density $\delta\rho({\bf r},t)$
in the response to a weak time-dependent external perturbation
$v_{ext}({\bf r},t)$, \beq \delta\rho({\bf r},t)=\int_{t_0}^{t}
\chi({\bf r},t;{\bf r}^{\prime},t^{\prime}) v_{ext}({\bf
r}^{\prime},t^{\prime})d{\bf r}^{\prime}dt^{\prime}. \eeq As a
consequence of eq. (2), $\chi$ have to satisfy the causality
condition: \beq \chi({\bf r},t;{\bf r}^{\prime},t^{\prime})=0
 \,\,\,\,\, if \,\,\, t^{\prime}>t. \eeq Therefore, one can assume,
that the following equality, \beq \chi^{-1}
({\bf r},t;{\bf r}^{\prime},t^{\prime})=0 \,\,\,\,\, if \,\,\,
t^{\prime}>t,\eeq is also always correct. By comparing eq.  (4) and
eq. (1) a conclusion can be made that there exists a paradox, since
the causality and symmetry contradict each other.  In their paper,
Harbola and Banerjee gave arguments that eq. (4) is not valid and the
inverse linear response function $\chi^{-1}$ can be only a symmetric
function of its time variables.  Thus, they conclude that there is no
conflict between the symmetry requirement and the causality in
time-dependent density functional theory \cite{hb}.

In this Comment we show that the statement, that the inverse of
the linear response function  $\chi^{-1}$ has to be only a
symmetric function of its time variables, is incorrect. In fact,
the eq. (1) has a solution which preserves the causality. It means
that an external field $v_{ext}$ can be defined by the density
$\rho$ in the form being similar to eq. (2). Then, we will briefly
discuss the interplay between the causality and response
functions.

For the sake of simplicity and in order to concentrate on the time
variables, we choose external field of the form
$v_{ext}({\bf r},t)=\hat{q}({\bf r})v_{ext}(t)$,
with $\hat{q}({\bf r})$
being the operator of a physical quantity.
The transition to the general case is straightforward.
Inserting this expression for the external field into eq. (2) and
omitting for simplicity the spatial variables, one gets,
\beq \delta\rho(t)=\int_{t_0}^{t} \chi_{\tau}(t,t^{\prime})
v_{ext}(t^{\prime}) dt^{\prime}, \eeq where
$\chi_{\tau}(t,t^{\prime})$ is the time-dependent part of the
function $\chi$. It is seen that eq. (5) is the Volterra integral
equation of the first kind, which can be reduced to the equation
of the second kind by differentiation with respect to $t$. This
leads  \cite{vt} to the following equation:
\beq \frac{1}{K(t)}
\frac{\partial^2\delta\rho(t)}{\partial t^2}=v_{ext}(t)
+\int_{t_0}^{t} \frac{1}{K(t)}\frac{\partial
^2\chi_{\tau}(t,t^{\prime})}{\partial t^2}v_{ext}(t^{\prime})
dt^{\prime}, \eeq where $K(t)$ is given by the following
expression: \beq K(t)=\frac{\partial\chi_{\tau}(t,t^{\prime})}
{\partial t^{\prime}} |_{t=t^{\prime}}.\eeq Eq. (6) always has a
solution \cite{vt}, \beq v_{ext}(t)=\frac{1}{K(t)}
\frac{\partial^2\delta\rho(t)}{\partial t^2}+ \int_{t_0}^{t}
R(t,t^{\prime}) \frac{1}{K(t^{\prime})}
\frac{\partial^2\delta\rho(t^{\prime})} {\partial t^{\prime
2}}dt^{\prime}, \eeq with $R(t,t^{\prime})$ being the resolvent,
which meets the condition, \beq R(t,t^{\prime})=0 \,\,\,\,\, if
\,\,\, t^{\prime}>t.\eeq As it is seen from eqs. (8) and (9) the
dependence of the potential $v_{ext}({\bf r},t)$ upon the density
$\rho({\bf r}^{\prime},t^{\prime})$ may be restricted to the times
$t^{\prime}<t$.
We remark, that the derivative on the right of eq. (8) may be
written as,
\beq
\frac{\partial f(t^{\prime})}{\partial t^{\prime}}=
\frac{f(t^{\prime}+h)-f(t^{\prime})}{h}|_{h\to -0},
\eeq
with $f(t^{\prime})$ presenting $\delta\rho(t^{\prime})$ or
$\partial\delta\rho(t^{\prime})/\partial t^{\prime}$.
It is seen from eq. (10) that the derivative can be calculated
sampling $t^{\prime}\leq t$.
Thus, the solution of eq. (5) can be presented in
the form,
\beq v_{ext}(t)=\int_{t_0}^{t} \chi_{\tau}^{-1}(t,t^{\prime})
\delta\rho(t^{\prime}) dt^{\prime}, \eeq
with $\chi^{-1}$ satisfying eq. (4).
This result is contradictory to the conclusion of
\cite{hb} that $\chi^{-1}$ has to be only a symmetric function of
its time variables. The contradiction results from the incorrect
assumption of \cite{hb} that the inverse $\chi^{-1}$ may be
proportional to $\delta(t-t^{\prime})$ at most, while, as it is seen
from eqs. (8,11), the inverse is proportional to the second
derivative of $\delta$-function with respect to the time variable.
On the other hand, in special situations, our result very likely
lends support to the main result of \cite{hb}, showing that the
causal inverse incorporates differential operator (10).
Our consideration was restricted to the case of weak external
perturbation $v_{ext}({\bf r},t)$. In the event of strong external
fields, changing, for instance, suddenly at time $t$, the derivative
on the right of eq. (8) cannot be defined by eq. (10), and it is
probably impossible to construct the causal inverse.

Thus, we would like to emphasize, that at least in the case of
weak external fields, there exist both the causal inverse of linear
response function $\chi$, and the symmetric (noncausal) inverse
$\chi^{-1}$ \cite{hb,ksk,as}, that is equality (4) is not always
correct. Moreover, one can show that it is also possible to
construct the advanced inverse.  All these functions can be used to
construct the causal response function $\chi$ satisfying eq. (3). But
eq. (1) defines the symmetric inverse $\chi^{-1}$, which is to be
used without any violations of the causality \cite{hb}. As a result,
it is the existence of noncausal $\chi^{-1}$ that leaves no room for
the paradox. This result has been obtained \cite{ksk} in fact long
before the paradox was invented \cite{gdp,bg,l}, and the conclusions
of \cite{ksk} were confirmed in \cite{hb}. It is worth to mention,
that a close examination of the relations between the action
integral, response functions, effective interaction (or
exchange-correlation kernel), single-particle Kohn-Sham
time-dependent potential, and causality in the time-dependent density
functional theory has been carried out recently in \cite{as}.

This research was funded in part by INTAS under Grant No.
INTAS-OPEN 97-603. We also wish to thank M.K. Harbola for
useful discussions.

\end{document}